# Fabrication of Graphene *p-n-p* Junctions with Contactless Top Gates

Gang Liu, Jairo Velasco Jr., Wenzhong Bao and Chun Ning Lau[*]

Department of Physics & Astronomy, University of California, Riverside, CA 92521

We developed a multi-level lithography process to fabricate graphene *p-n-p* junctions with contactless, suspended top gates. This fabrication procedure minimizes damage or doping to the single atomic layer, which is only exposed to conventional resists and developers. The process does not require special equipment for depositing gate dielectrics or releasing sacrificial layers, and is compatible with annealing procedures that improve device mobility. Using this technique, we fabricate graphene devices with suspended local top gates, where the creation of high quality graphene *p-n-p* junctions is confirmed by transport data at zero and high magnetic fields.

---

[*] To whom correspondence should be addressed. E-mail address: lau@physics.ucr.edu.

Graphene, a single-layer honeycomb lattice of carbon atoms, has recently emerged as a fascinating system for fundamental studies in condensed-matter physics[1-4]. Because of its unusual band structure, single-layer graphene is a zero-gap semiconductor with a linear energy dispersion relation, where charges behave as massless Dirac fermions. Charge transport in graphene exhibit a number of novel phenomena, such as the half-integer quantum Hall effect[1,3] and bipolar supercurrent transistors. Technologically, graphene is a two-dimensional material with exceptional mobility, current-carrying capacity and thermal conductivity, attracting significant attention as a promising post-silicon electronic material[5-12].

A remarkable electronic property of graphene is that both carrier type and density can be controlled electrostatically. Via the employment of a local gate and a global back gate, this feature also enables *in situ* creation and control of *p-n* junctions in graphene[13,14], which have been demonstrated or predicted to give rise to quantum Hall plateaus with fractional values due to mode mixing of edge states[16-18], Vaselago lensing[19,20] and Klein tunneling[21]. In most of the experiments to date, either an organic or a metal oxide layer has been used as the local gate dielectric, yet its deposition on a single atomic layer remains a delicate process that may lead to additional dopants and/or scattering sites. Moreover, dramatically enhanced mobility has been recently observed in suspended graphene devices[22,23], but fabrication of suspended graphene *p-n* junctions using conventional techniques could prove difficult, since direct deposition of local gate dielectrics may considerably stress or even collapse the atomic layer.

Here we report an innovative multi-level lithography technique to fabricate contactless top gates that are suspended ~100 nm above the graphene layers, which are only exposed to conventional electron-beam resists and developers, thus minimizing the damage to the atomic layer. The vacuum insulated gap between the "air bridge"-styled top gate and the device is not susceptible to pinholes, dielectric breakdown and leakage current. Unlike the standard graphene *p-n* junctions, devices fabricated using this technique are still amiable to annealing procedures that have been shown to be effective in improving device mobility and contact resistance[22,24].Using this technique, we demonstrate the fabrication of a graphene *p-n-p* junction, and its quality is established by the presence of the $2e^2/h$ conductance plateau at high magnetic fields. In the long term, combination of this technique with suspended graphene may enable experimental realization of novel phenomena such as the Veselago lensing effect.

This fabrication process of suspended structures takes advantage of the different exposure, developing and lift-off properties of different resists. Fig. 1 illustrates the fabrication process, which consists of two lithography steps and a single metal deposition step. The goal of the first lithography step is two-fold: (1) to deposit a resist layer that will act as an initial mechanical support for the suspended portion of the air bridge, and will be removed at the end of the fabrication; and (2) to create windows for the electrical leads that contact the suspended structure. To achieve these goals, LOR/PMMA bilayer resists are spun and baked on Si/SiO$_2$ substrates (Fig. 1a)[25]. We then use an electron beam to expose the patterns for the electrical leads. The chips are developed twice – first in MIBK/IPA to dissolve the exposed PMMA, then in MF319 to dissolve portions of LOR via the PMMA windows (Fig. 1b). Subsequently, acetone is used to lift off the top PMMA layer while leaving LOR intact (Fig. 1c). The final outcome of the first lithography step is an LOR layer with windows for electrical contacts for the air bridge.

During the second lithography step, MMA/PMMA bilayer resists are spun and baked on top of the LOR layer, followed by electron beam exposure of patterns for both the air bridge and the electrical leads (Fig. 1d). The chips are developed in MIBK that removes exposed MMA and PMMA, leaving windows in MMA/PMMA bilayer for the air bridge, and windows in all three resist layers for the leads (Fig. 1e). Lastly, the device is completed by two metal depositions at 45° and -45°(Fig. 1f), so as to deposit metals onto the side-walls of the windows and ensure contact between the suspended structure and the electrical leads. For the final lift-off process, the 3 resist layers are removed by PG remover, (Fig. 1g), and the chips are rinsed in isopropyl alcohol and dried in nitrogen gas.

Examples of completed suspended structures are shown in Fig. 2. The fabrication procedure is quite robust; by controlling the lithography conditions, we are able to fabricate suspended air bridges with considerable ranges in dimensions, including span ($l$), width ($w$) and height above the substrate ($h$). For our purpose of using a suspended bridge as a local top gate, the important parameters are $l$ and $h$: the former directly limits the width of the graphene strip that can be used, the latter determines the gate efficiency. The height of the bridge is determined by the thickness of the LOR resist layer, which may range from 50 nm to 3 μm. Fig. 2a and 2b display two bridges that are suspended 300 nm and 100 nm above the substrates, respectively. On the other hand, we find that $l$ increases with $w$ and the material's strength. A few bridges with different $w$ are shown in Fig. 2c: from top to bottom, the bridges are 250, 200, 150 and 100 nm wide, respectively. Clearly, the 100 nm-wide bridge sags in the center, while the 250 nm-wide bridge remains straight. We note that we are able to create titanium air bridges ~ 7 μm long *without critical point drying*. Such bridges, or similarly suspended structures, can be used for a number of applications such as local injection of current and nanoelectromechanical devices.

To demonstrate an application of this technique, we fabricate graphene junctions with local top gates (Fig. 3). A completed *p-n-p* device, similar to that shown in Fig. 3c, is measured at 260mK using standard lock-in techniques. The device's source-drain separation is 3.5 μm, with a top gate that covers a ~0.5 μm-long segment in the center and is suspended ~100 nm above the substrate.

The device's differential resistance $R$ is plotted in Fig. 4a as functions of the back gate voltage $V_{bg}$ (vertical axis) and top gate voltage $V_{tg}$ (horizontal axis). The most visible feature is the red horizontal band at $V_{bg}$ ~ 14V, corresponding to the Dirac point of the entire graphene sheet between the source and drain electrodes. The device mobility is estimated to be ~8500 cm$^2$/Vs. Another notable feature is the diagonal white band, indicating the Dirac point for the top-gated region. The presence of two Dirac points is more easily seen in Fig. 4b, which plots $R$ vs $V_{bg}$ at three fixed $V_{tg}$, corresponding to cuts along the dotted lines in Fig. 4a. Apart from the prominent center peak, the orange curve has an additional shoulder at $V_{bg}$~-14V, and the blue curve at $V_{bg}$~28V, corresponding to the Dirac point of the top-gated portion. We note that shoulders, rather than full-blown side peaks, are observed at the second Dirac point; this is because the top gate controls less than 15% of the entire device's area. Our results clearly demonstrate individual control of separate regions in the graphene device, and the formation of a graphene *p-n-p* junction.

The slope of the white line in Fig. 4a yields the ratio of the coupling efficiencies $\eta$ of the two gates to graphene, and is determined to be $\frac{\eta_{bg}}{\eta_{tg}} = \frac{\Delta V_{tg}}{\Delta V_{bg}} \sim 1.27$. From simple geometry consideration, $\eta$ is given by the gate-device capacitance per unit area, $C=\varepsilon\varepsilon_0/d$, where $\varepsilon$ is the dielectric constant of the gate dielectric (3.9 for SiO$_2$), $\varepsilon_0$ is the permittivity of free space, and $d$ is the gate-device separation. Hence, the coupling ratio is given by $\frac{C_{bg}}{C_{tg}} = \frac{\varepsilon_{bg}}{\varepsilon_{tg}} \frac{d_{tg}}{d_{bg}} \approx$ (3.9)(100/300)≈1.3, in excellent agreement with the data. Furthermore, the typical voltage range that the top gates can sustain is ~50-75V; we expect this value to increase as we optimize the fabrication process, *e.g.* by improving the contacts between the vertical walls and the suspended bridge.

Further evidence for the formation of graphene *p-n-p* junctions is provided by transport data at magnetic field of 8T. For a graphene device with uniform carrier density in high magnetic fields, the formation of quantum Hall edge states leads to a series of conductance plateaus at half integer values of $4e^2/h$ (the factor of 4 originates from the spin and valley degeneracy). This so-called "half-integer quantum hall effect" arises from the linear dispersion relation of single-layer graphene, and can be attributed to the presence of a Landau level at zero energy, shared by both electrons and holes. In graphene *p-n-p* junctions, we observe additional conductance plateaus at fractional values (such as 6/5 and 2/3) of $e^2/h$ (Fig. 4c-d), similar to those reported by Ozyilmaz *et al.*[18]. As shown in Fig. 4e, such fractional plateau values arise from the partial and full equilibration of the edge states at the *p-n* interfaces[18], depending on the signs and magnitudes of $v$ and $v'$, the filling factors in the uncovered and top-gated regions, respectively. The values of $v'$ (at $v$=0) and $v$ are labeled at corresponding gate voltages in Fig. 4d. In particular, we note that the conductance plateau with the full value of $2e^2/h$ was not observed in ref. 18, due to the strong effect of backscattering on states with $|\nu|=|\nu'|$. In contrast, we observed a clear plateau at $2e^2/h$ (Fig. 4d), establishing that transport in our *p-n-p* junctions experiences relatively weak backscattering.

In conclusion, we have developed a multi-level lithography process to fabricate suspended top gates ranging from sub-100nm to a few μm in size. Graphene *p-n-p* junctions with such suspended gates exhibit high mobility and local control of doping density and type; observation of the previously unreported $2e^2/h$ quantum Hall plateau in similar *p-n-p* junctions demonstrates that our procedure produces clean junctions. In the long term, this versatile technique can also be significantly improved, and extended to fabricate other types of suspended structures such as moving parts in micro-electro-mechanical devices.

We thank Marc Bockrath for helpful discussions, and Peng Wei for assistance with data acquisition software. The research was supported in part by NSF CAREER grant no. DMR/0748910 and ONR/DMEA Award H94003-07-2-0703.

**Figure Captions**

**Figure 1.** Schematics of fabrication process. (a). LOR(blue) and PMMA(brown) are deposited onto the substrate, and exposed to electron beams (arrows). (b) Developing in MIBK and MIF319 solutions opens windows for the electrodes. (c). Liftoff in acetone removes PMMA but leaves LOR layer intact. (d). MMA(green) and PMMA(brown) are deposited, and exposed to e-beam (arrows). (e). Developing in MIBK opens windows for the electrodes and the suspended structure. (f). Metals are evaporated at 45º and -45º (in directions indicated by arrows in e). (g). Resists are lifted off in PG remover, leaving an air bridge contacted to electrodes.

**Figure. 2**. SEM images of suspended air bridges with different span, width and height. Scale bars: 2 μm.

**Figure 3.** SEM images of graphene devices with suspended air bridges. (a) (c) top view. (b) angled view (60º) of the device in (a). Scale bars: 2 μm.

**Figure 4.** (a) Differential resistance of a graphene device with a center top gate covering ~ 15% of the device area, as functions of $V_{bg}$ and $V_{tg}$. The dotted lines correspond to the line traces in (b). Doping combinations for different regions are labeled. (b) Differential resistance as a function of $V_{bg}$ at different $V_{tg}$. (c). Device conductance at 8T magnetic field, as functions of $V_{bg}$ and $V_{tg}$. For reference, the filling factors $\nu$ and $\nu'$ are also labeled (the latter at $\nu=0$). They are related to the gate voltages by $n_c h/eB$, where $h$ is the Planck's constant, $e$ is the electron charge, and $n_c$ is the charge density, given by $C_{bg}V_{bg}/e$ in the uncovered region, and $(C_{bg}V_{bg}+C_{tg}V_{tg})/e$ in the top-gated region. (d). Line trace along the red line in (c). The bracketed numbers correspond to $(\nu, \nu')$ for the plateaus. (e). Schematics of edge state propagation for different values of $(\nu, \nu')$.

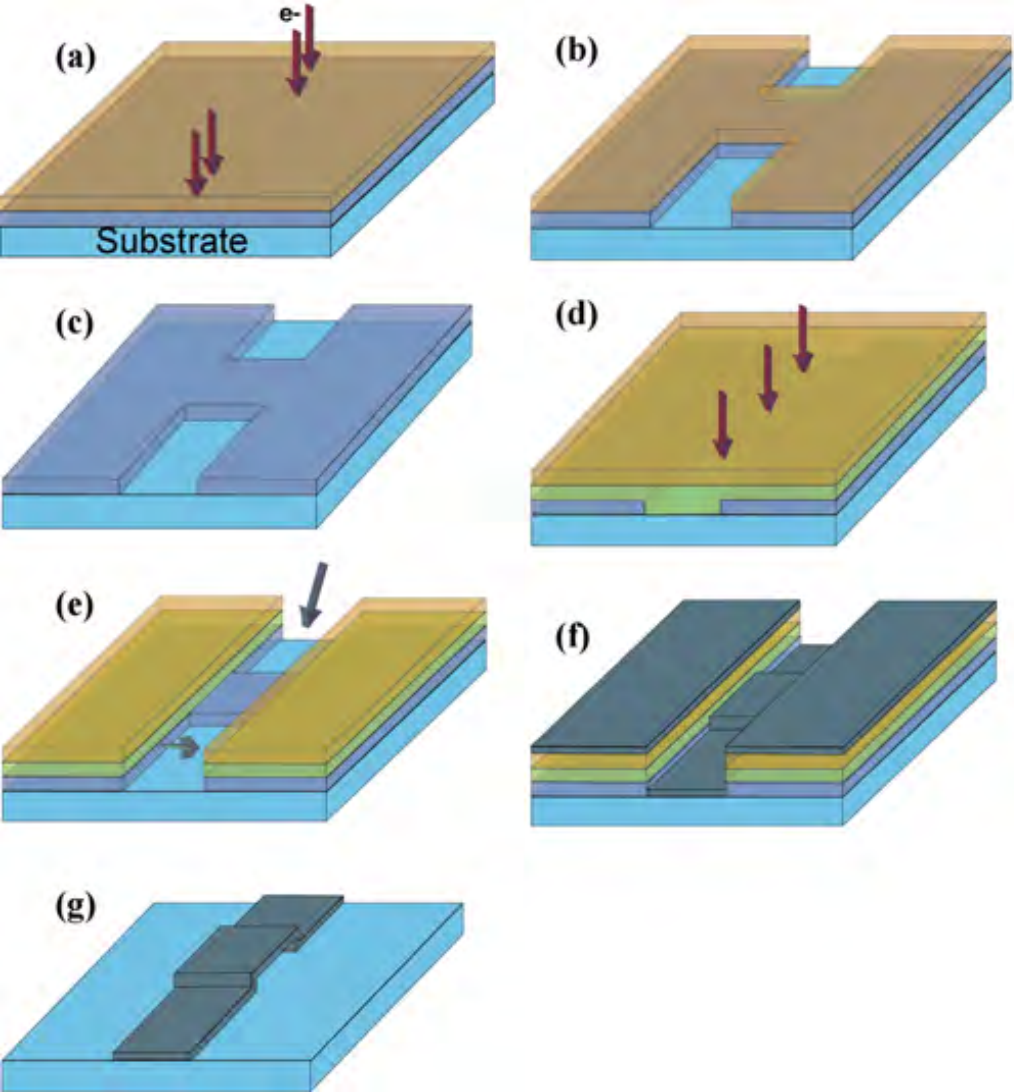

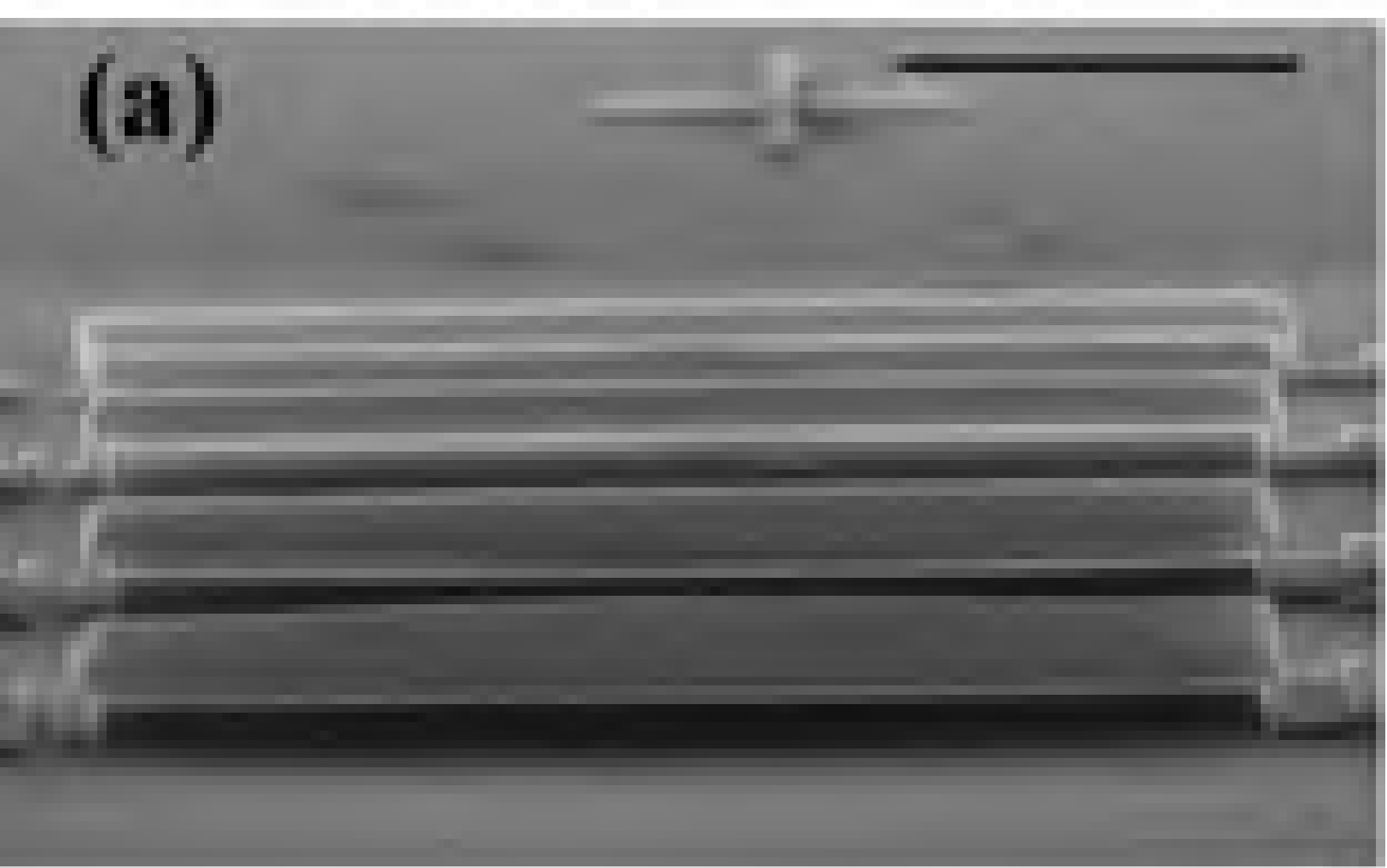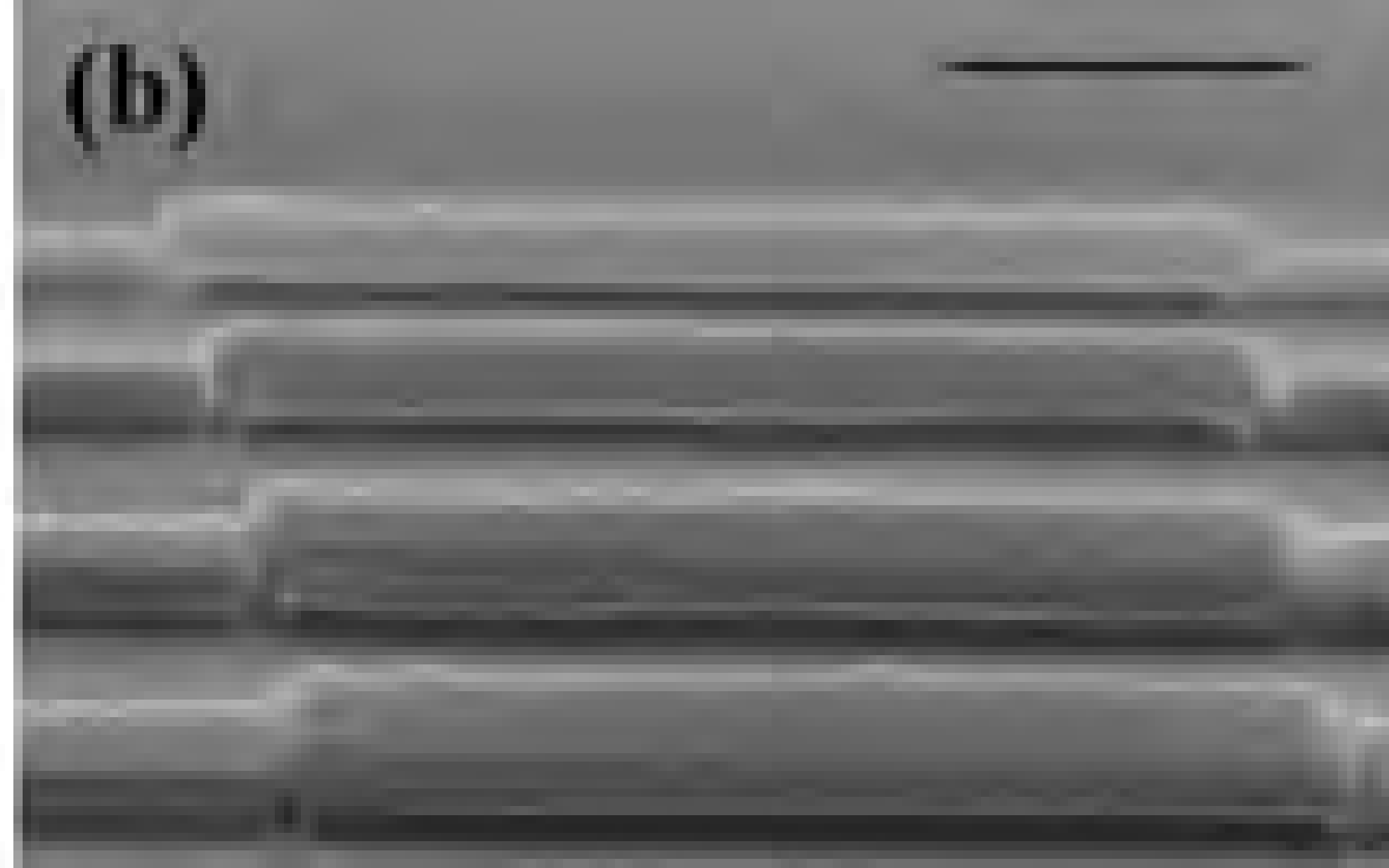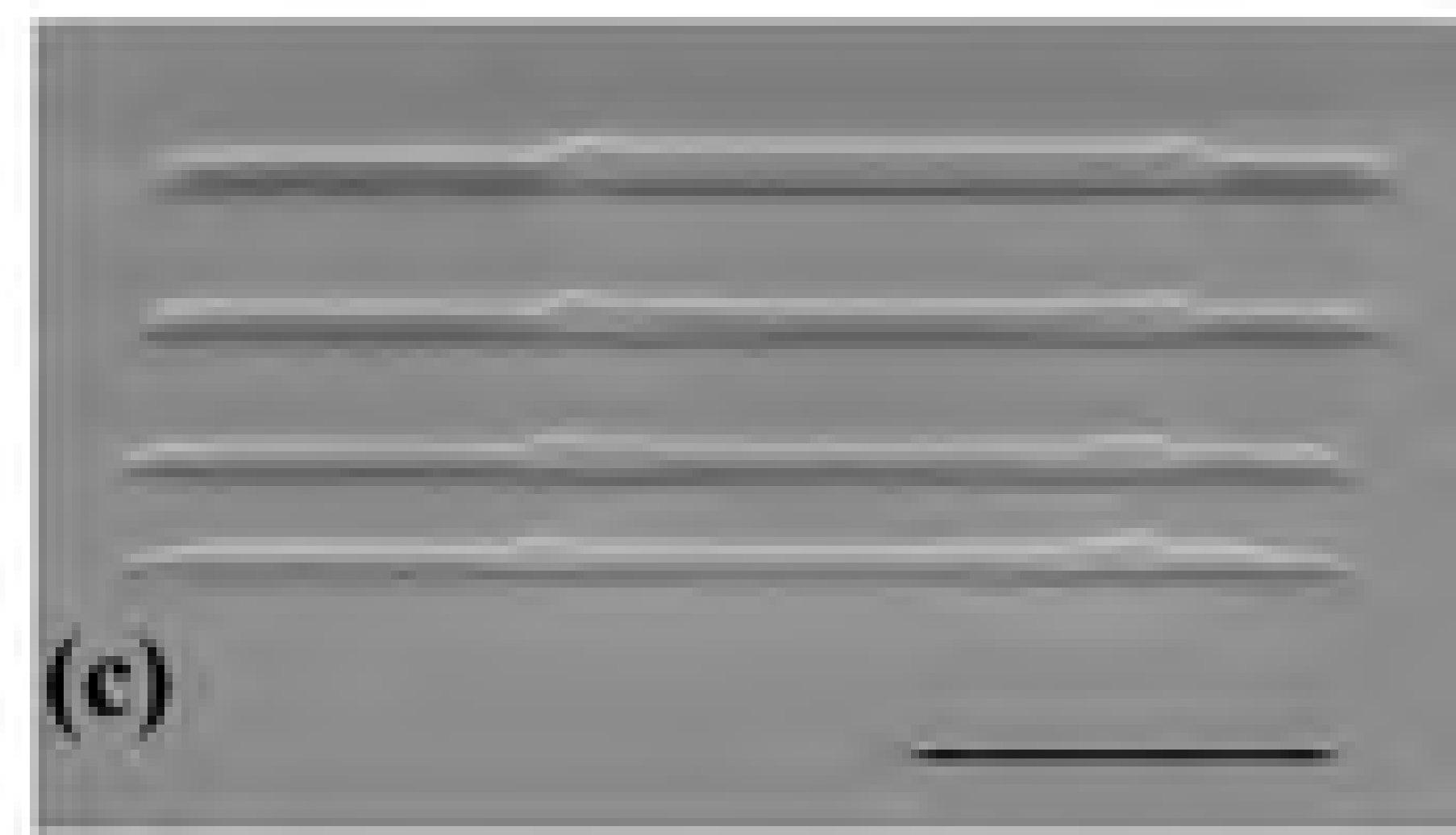

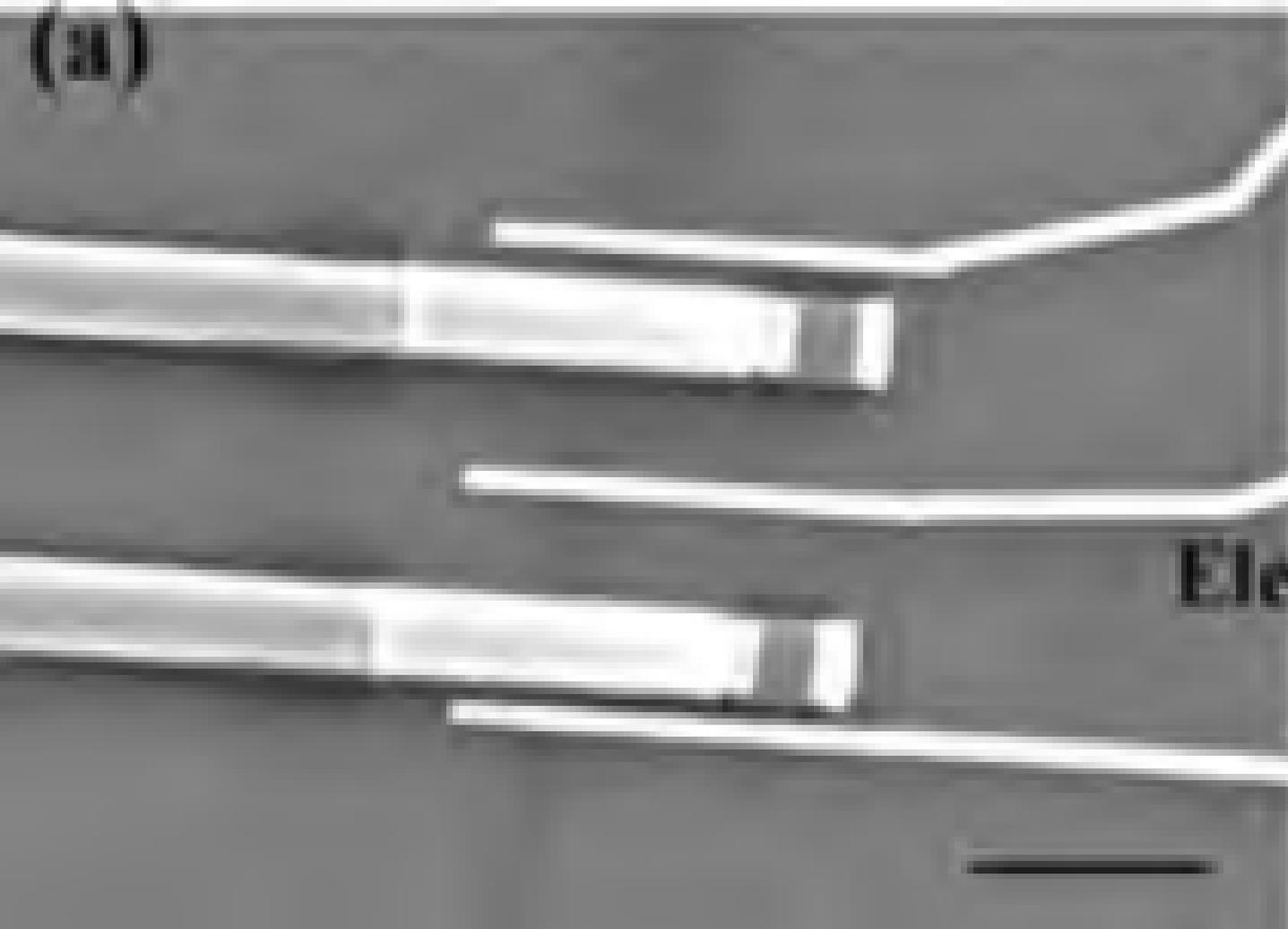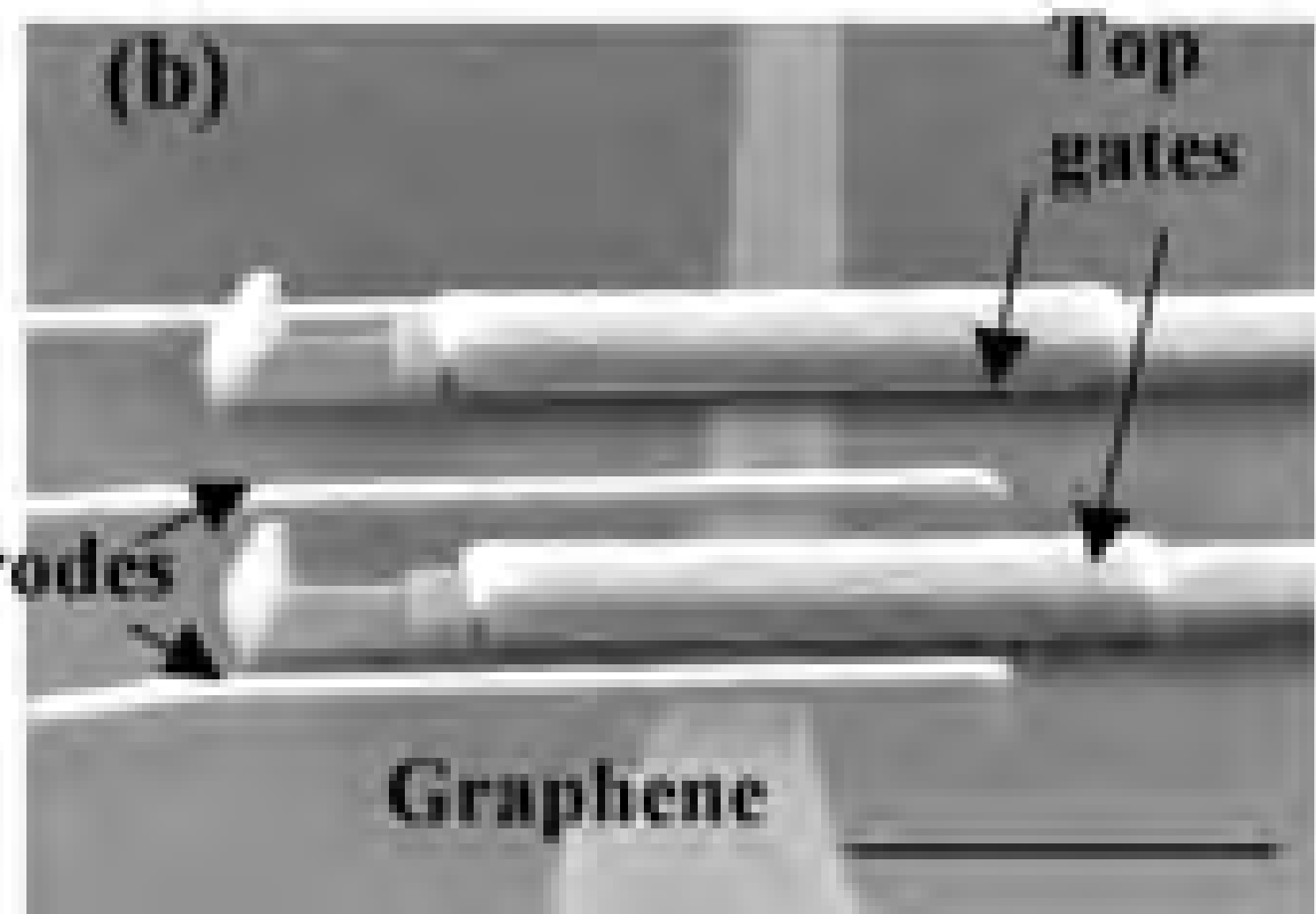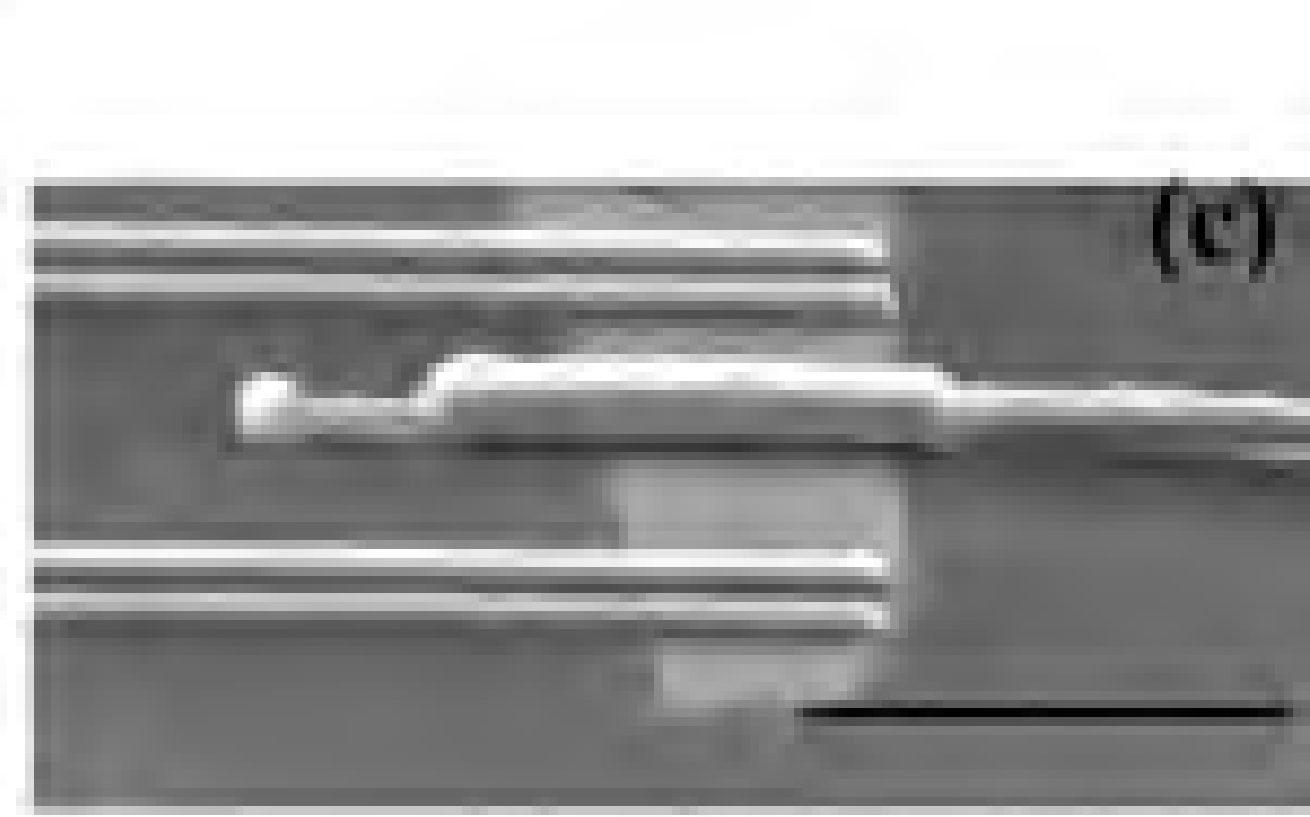

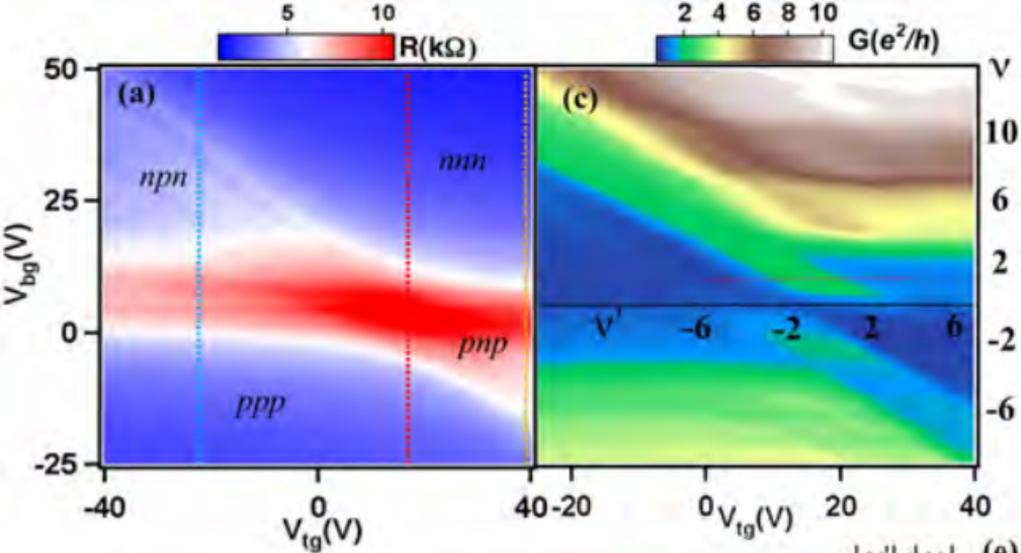

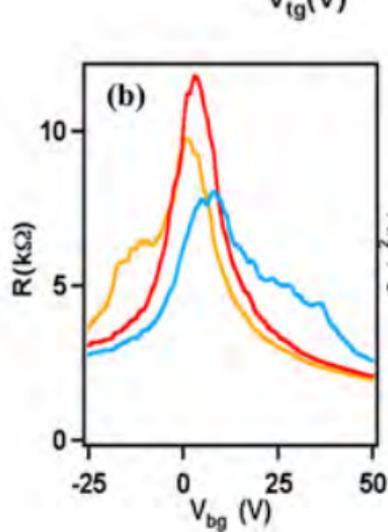
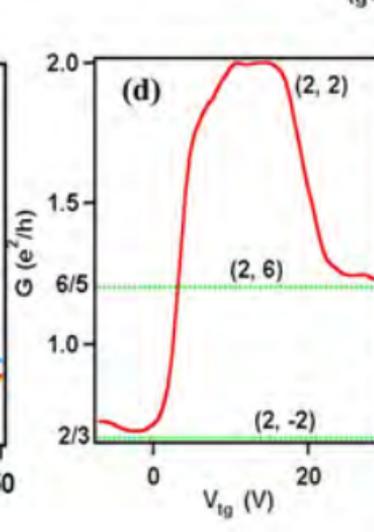
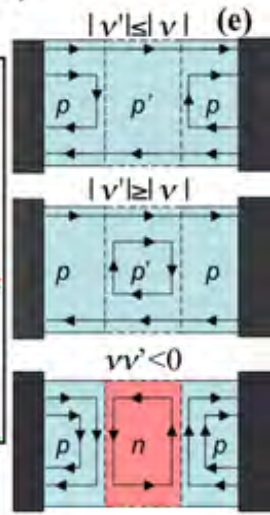